\documentclass[a4paper]{jpconf}
\usepackage{graphicx}
\usepackage[table,xcdraw]{xcolor}
\usepackage[frozencache,cachedir=.]{minted}
\newcommand\Mark[1]{\textsuperscript{#1}}
\begin{document}
\title{Efficient and Accurate Automatic
Python Bindings with cppyy \&
Cling}

\author{Baidyanath Kundu\Mark{$\ast$}\Mark{$\dagger$}, Vassil Vassilev\Mark{$\ast$}\Mark{$\dagger$}, Wim Lavrijsen\Mark{$\S$}}

\address{\Mark{$\ast$}European Council for Nuclear Research, Espl. des Particules 1, 1211 Meyrin, Switzerland \\ \Mark{$\dagger$}Department of Physics, Princeton University, Princeton, New Jersey 08544, USA \\ \Mark{$\S$}LBNL, MS 50B-6222, 1 Cyclotron Road, Berkeley CA 94720-8147, USA}

\ead{baidyanath.kundu@cern.ch, vassil.vassilev@cern.ch, wlavrijsen@lbl.gov}

\begin{abstract}
The simplicity of Python and the power of C++ force stark choices on a scientific software stack. There have been multiple developments to mitigate language boundaries by implementing language bindings, but the impedance mismatch between the static nature of C++ and the dynamic one of Python hinders their implementation; examples include the use of user-defined Python types with templated C++ and advanced memory management.

The development of the C++ interpreter Cling has changed the way we can think of language bindings as it provides an incremental compilation infrastructure available at runtime.
That is, Python can interrogate C++ on demand, and bindings can be lazily constructed at runtime. This automatic binding provision requires no direct support from library authors and offers better performance than alternative solutions, such as PyBind11. ROOT pioneered this approach with PyROOT, which was later enhanced with its successor, cppyy. However, until now, cppyy relied on the reflection layer of ROOT, which is limited in terms of provided features and performance.

This paper presents the next step for language interoperability with cppyy, enabling research into uniform cross-language execution environments and boosting optimization opportunities across language boundaries. We illustrate the use of advanced C++ in Numba-accelerated Python through cppyy. We outline a path forward for re-engineering parts of cppyy to use upstream LLVM components to improve performance and sustainability. We demonstrate cppyy purely based on a C++ reflection library, \textit{InterOp}, which offers interoperability primitives based on Cling and Clang-Repl.
 
\end{abstract}

\section{Introduction}

The C++ programming language was adopted in the mid-`90s by the high energy physics (HEP) community to provide a more modern yet performant programming environment. At the time, HEP had some of the world's largest data sets (and processing rates). Python usage in HEP started in the mid-2000s, initially only for interactive access to experiments’ frameworks and analysis. Meanwhile, the rest of the world caught up with and even surpassed HEP data processing needs, and a complete ecosystem has grown up around Python to support data science. Much of the modern software for HEP physics analysis, such as Awkward Array~\cite{AwkwardArray}, Uproot~\cite{Uproot}, and Coffea~\cite{Coffea}, is grounded in this Python ecosystem. C++ remains favored because of its performance, access to accelerators in heterogeneous computing environments, and having foundational libraries in HEP, such as ROOT~\cite{ROOT}, GEANT4~\cite{GEANT4}, and most experiments’ processing  and production software frameworks. With both languages playing important roles, Python-C++ integration has become instrumental.
However, more advanced integration is necessary for high-performance codes and codes that run in heterogeneous environments. 
For example, a data processing task performed by a C++ framework running a user-provided Python function or kernel incurs significant overhead because of the large number of language crossings.
Similarly, if the task is run as part of a GPU workflow, it will incur a performance penalty due to offloading since the Python code can only run on the CPU.

Numba~\cite{Numba}, a just-in-time compiler (JIT) for Python code, addresses this problem: it is capable of compiling Python code, targeting either the CPU or GPU, and provides callable interfaces to use the JITed closures from low-level libraries. Thus, Numba greatly improves the performance of Python code not only by lowering it to machine code but also by removing unnecessary crossings of language (or device) boundaries in inner loops. However, Numba has two shortcomings: preparing external code, such as functions written in C, to be usable for the Numba JIT is an involved, manual process; and Numba does not support C++ directly.

Cppyy, an automatic runtime bindings generator~\cite{cppyy} based on the C++ interpreter Cling~\cite{Cling}, can address both these shortcomings.
One goal of our work is to bring advanced C++, in particular highly optimized numerics libraries, into Numba-accelerated Python in a fully automatic way, with minimal language crossing overhead.
In this paper, we present a fully generic prototype that provides, lazily and at runtime, Numba extensions for bound C++ through cppyy.
Our prototype supports overloading, C++ templates, data member access, and instance method calls in Numba-accelerated Python.
We outline a path forward for re-engineering cppyy's backend to use LLVM components more directly, thus improving performance and sustainability.
We demonstrate cppyy on InterOp, a new library that implements interoperability primitives based on Cling and Clang-Repl.

This paper is organized as follows: Section~\ref{sec:motivation} motivates our technology and design choices; Section~\ref{sec:implementation} provides a bird’s eye overview of the implementation design; Section~\ref{sec:results} shows timing results with the current implementation; Section~\ref{sec:improvements} discusses the enhancements brought about by a new reflection library, InterOp, which will allow for even better performance and greater flexibility when retargeting codes; Section~\ref{sec:related} discusses related work; finally, Section~\ref{sec:conclusion} summarizes our findings.

\section{Motivation}
\label{sec:motivation}
The typical approach to speed up a Python application has been to write the performance-critical portions of it in a lower-level language, usually C, and then access that code in the Python interpreter through an extension module. A much more elegant solution is to lower the original Python code to native through the use of a JIT so that the developer can stay within a single environment, Python, and only needs to write and debug the code once. In HEP, it is critical that such a JIT works well with bound C++ code and that JITed Python code is usable from C++, since any full HEP software stack relies heavily on both languages. For this paper, we use Numba as the Python JIT and integrate C++ into it through cppyy.

Numba has an extension application programming interface (API) to add functions and types for the JIT to recognize, annotate, and lower to LLVM IR (intermediate representation; which in turn gets assembled to native code). Manually writing such extensions is tedious and time-consuming, so automation is critically important, especially to meet the goal of being able to use the JIT transparently. Integrating C++ with Numba would additionally make it easy to use Python kernels in C++ code without losing performance, thus reaping the benefits of continued use of a large installed base of C++ codes.
To enable this, we use cppyy to generate first-class Python objects for bound C++, which enables Numba to query for reflection information through regular Python introspection and for cppyy to provide the necessary Numba extensions lazily at runtime. The result is a generic, compact, and efficient implementation.

Modern C++ is especially amenable to automation: successive C++ standards have moved the language away from low-level code to more expressive abstractions for the purposes of code quality and optimizing compilers. In modern C++, many previously difficult to analyze corner cases are now more clear; compare, e.g., ownership rules for raw v.s.\ smart pointers. Furthermore, cppyy is based on Cling, the C++ interpreter, to have C++ match Python’s dynamism, interactivity, and runtime behavior. Cling itself is based on Clang/LLVM, which allows it to stay up to date even as the C++ language continues to evolve rapidly. The use of Cling also means that unresolved corner cases can be handled at runtime in either C++ or Python as appropriate, without necessitating an intermediate bridging or mapping language.


\section{Implementation}
\label{sec:implementation}
We introduce C++ into Numba through a new reflection interface on top of cppyy, which resolves overloaded functions and instantiates C++ templates during Numba’s type annotation step; and which provides the low-level description for lowering to LLVM IR for subsequent optimization and conversion of assembly to machine code. This process is completely automatic and transparent, including integration with other manually provided Numba extensions.
Performance is on par, with only a moderate additional warm-up cost.
Python programmers can thus continue to develop and debug their codes fully in Python while simply switching on the Numba JIT for selected performance-critical closures. 

Python is dynamically typed.
It achieves this by wrapping objects into so-called PyObjects via a method known as \textit{boxing} and retrieves their original value via a method known as \textit{unboxing}. Since Numba needs to eliminate these Python \textit{boxing} and \textit{unboxing} steps, it requires accurate typing information ahead of time.
To provide detailed, exact and low-level C++ type information to Numba, which is needed in addition to the type information already available through Python's introspection, we introduce a new \textit{Reflection API} to cppyy.

The reflection API follows the request/reply pattern and adds a new function to all cppyy-bound  objects called \mintinline{python}{__cpp_reflex__}. The function has a constant signature for all cppyy object types, but its functionality depends on which one it is called on. It takes two arguments, the type of reflection information and its format, and returns the information requested. The request can range from asking if the object represents a namespace or class to requesting the object's C++ type. The format argument specifies whether we want the information in a string format, a C++ type, or in the manner that the API thinks is the most optimal.

\begin{figure*}[h]
    \centering
    \includegraphics[width=0.9\textwidth]{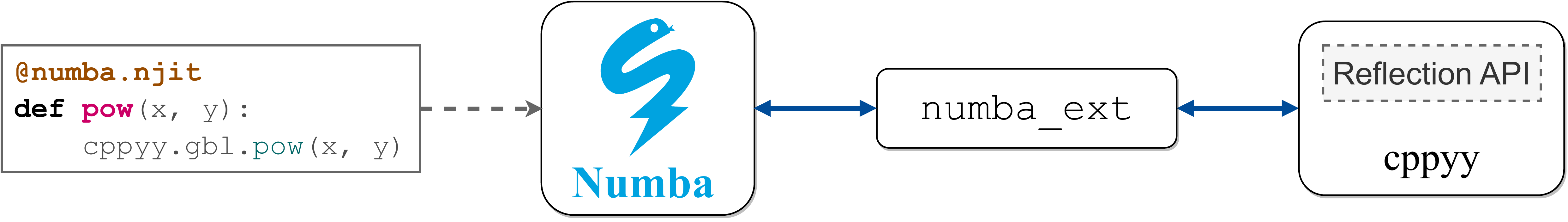}
    \caption{The interaction diagram for Numba, numba\_ext and cppyy}
    \label{fig:numba_pipeline}
\end{figure*}

Fig.~\ref{fig:numba_pipeline} shows the layered approach taken by Numba to separate concerns.
Numba analyzes a Python closure as directed, and upon encountering cppyy types, it queries cppyy's pre-registered \textit{numba\_ext} module for exact type information.
This module, in turn, queries cppyy's new reflection API if it encounters a type that has not been seen before.
The module then uses Numba's extension and core modules to generate the necessary typing classes, to be returned to Numba during this typing step; and lowering methods, which Numba will expect in its next step, to write the LLVM IR.
Each core language construct (free functions, namespaces, classes, methods, data members, etc.) has its own implementation.
For instance, to support C++ free functions, we extend the Numba \textit{Callable} type and register it as the type provider for cppyy function objects.
When Numba encounters a C++ function bound by cppyy, it queries this registered type for the function signature (i.e., the return and argument types), given the actual argument types traced so far.
Cppyy's \textit{numba\_ext} uses the argument types to select an appropriate overload, aided by the new reflection API, and wraps this overload in a unique typing class.
Finally, \textit{numba\_ext} registers a lowering method uniquely generated for the address of the chosen C++ function overload, which will be called during the Numba lowering step. This process provides Numba with all the information it needs to convert the function call to LLVM IR.

\section {Results}
\label{sec:results}
We benchmark the running time of Numba JITed functions with cppyy objects against their Python counterparts to obtain the time taken by Numba to JIT the function (Numba JIT time), the time taken by cppyy to create the typing info and possibly perform lookups and instantiate templates (cppyy JIT time), the time taken to run the function after it has been JITed (Hot run time), and the time taken to run the equivalent Python function. The results are obtained on a 3.1GHz Intel NUC Core i7-8809G CPU with 32G RAM.

\begin{figure*}[h]
\definecolor{higlightcolor}{HTML}{9AFF99}
\centering
\begin{minipage}[t]{0.2\textwidth}
\begin{minted}
[
frame=none,
framerule=1pt,
framesep=0mm,
baselinestretch=1,
fontsize=\scriptsize,
]
{python}

cppyy.cppdef(r"""
template<class T>
T add42(T t) {
    return T(t+42);
}
""")

\end{minted}
\centering{\footnotesize C++ code in cppyy}
\end{minipage}%
\begin{minipage}[t]{0.4\textwidth}
\setlength{\fboxsep}{1pt} 
\begin{minted}
[
frame=leftline,
framerule=1pt,
framesep=1mm,
baselinestretch=1,
fontsize=\scriptsize,
escapeinside=||,
]
{python}
|\phantom{empty space}|
def go_slow(a):
    trace = 0.0
    for i in range(a.shape[0]):
        trace +=
          |\colorbox{higlightcolor}{cppyy.gbl.add42(a[i, i]) +   }|
          |\colorbox{higlightcolor}{cppyy.gbl.add42(int(a[i, i]))}|
    return a + trace
\end{minted}
\centering{\footnotesize Python/C++ binding with cppyy}
\end{minipage}%
\begin{minipage}[t]{0.4\textwidth}
\setlength{\fboxsep}{1pt} 
\begin{minted}
[
frame=leftline,
framerule=1pt,
framesep=1mm,
baselinestretch=1,
fontsize=\scriptsize,
escapeinside=||,
]
{python}
@numba.jit(nopython=True)
def go_fast(a):
    trace = 0.0
    for i in range(a.shape[0]):
        trace +=
          |\colorbox{higlightcolor}{cppyy.gbl.add42(a[i, i]) +   }|
          |\colorbox{higlightcolor}{cppyy.gbl.add42(int(a[i, i]))}|
    return a + trace
\end{minted}
\centering{\footnotesize Numba-acceleration of cppyy}
\end{minipage}

\caption{\textbf{Benchmark fixture for `Templated free functions' case.} On the left side, the C++ templated function is declared in cppyy. In the center, we use a Python kernel to run this C++ function. On the right, we use the same kernel but add the Numba JIT decorator to accelerate it and compare the timing of the two kernels.  The other cases use the same setup, with only the highlighted code being replaced by the case being benchmarked.}
\label{lst:fixture}
\end{figure*}

To evaluate the speedup obtained by Numba JITing of cppyy objects, we use the fixture in Fig.~\ref{lst:fixture}. For each benchmark case in Tab.~\ref{tab:numba_speedup}, a Numpy array of size $100 \times 100$ was passed to the function; times indicated are averages of 3000 runs.
Numba JITed functions achieve a minimum speedup of 2.3 times in the case of methods and a maximum speedup of nearly 21 times in the case of templated free functions.
Tab.~\ref{tab:numba_speedup} summarizes timings obtained on our chosen benchmarks.

\begin{table}[h]
\footnotesize
\centering
\begin{tabular}{l
>{\columncolor[HTML]{DAE8FC}}c 
>{\columncolor[HTML]{DAE8FC}}c 
>{\columncolor[HTML]{FFFFC7}}c 
>{\columncolor[HTML]{FFFFC7}}c 
>{\columncolor[HTML]{FFFFC7}}r }
\hline
\textbf{Benchmark Case}                                            & \multicolumn{1}{l}{\cellcolor[HTML]{DAE8FC}\textbf{\begin{tabular}[c]{@{}c@{}}Cppyy JIT\\ time (s)\end{tabular}}} & \multicolumn{1}{l}{\cellcolor[HTML]{DAE8FC}\textbf{\begin{tabular}[c]{@{}c@{}}Numba JIT\\ time (s)\end{tabular}}} & \multicolumn{1}{l}{\cellcolor[HTML]{FFFFC7}\textbf{\begin{tabular}[c]{@{}c@{}}Hot run\\ time (s)\end{tabular}}} & \multicolumn{1}{l}{\cellcolor[HTML]{FFFFC7}\textbf{\begin{tabular}[c]{@{}c@{}}Python run\\ time (s)\end{tabular}}} & \textbf{Speedup} \\ \hline
Function w/o args                                                  & 1.72e-03                                                                                                          & 3.33e-01                                                                                                          & 3.58e-06                                                                                                        & 1.73e-05                                                                                                           & $4.84\times$            \\
Overloaded functions                                               & 1.05e-03                                                                                                          & 1.35e-01                                                                                                          & 4.51e-06                                                                                                        & 3.47e-05                                                                                                           & $7.70\times$            \\
Templated free functions                                           & 8.92e-04                                                                                                          & 1.45e-01                                                                                                          & 3.48e-06                                                                                                        & 7.18e-05                                                                                                           & $20.66\times$           \\
Class data members                                                 & 1.43e-06                                                                                                          & 1.33e-01                                                                                                          & 5.87e-06                                                                                                        & 1.82e-05                                                                                                           & $3.10\times$            \\
Class methods                                                      & 2.16e-03                                                                                                          & 1.39e-01                                                                                                          & 6.06e-06                                                                                                        & 1.43e-05                                                                                                           & $2.36\times$            \\ \hline
\end{tabular}
\caption{\textbf{Performance comparison.} The columns \textbf{cppyy JIT time} and \textbf{Numba JIT time} represent respectively the time taken by cppyy to create the Python objects for C++ entities and time taken for Numba to compile the Python function. Both do not execute the compiled functions. Column \textbf{Hot run time} shows the time taken to run the Numba JITed function, whereas \textbf{Python run time} shows the time taken to run the uncompiled Python function with cppyy code. Column \textbf{Speedup} compares the \textit{Hot run time} to \textit{Python run time}.}
\label{tab:numba_speedup}
\end{table}
\vspace{-0.5cm}

\section{Further Improvements}
\label{sec:improvements}
A promising area of research is incorporating the LLVM IR of the C++ code directly into the Numba-generated LLVM IR, in effect inlining the C++ functions into Python.
This would open up more optimization opportunities for the compiler and simplify support for alternate targets, such as GPUs.
In practice, this requires IR to be sent from Cling to Numba or vice versa. One of the challenges here is that Numba uses the llvmlite~\cite{llvmlite} package to synthesize LLVM IR, while Cling operates directly on the underlying IR.
Since LLVM IR has no guarantees of backward or forward compatibility, this could make a rather fragile dependency.
Furthermore, the current design of Cling does not provide the right abstractions and does not have the necessary low-level infrastructure to either extract or inject IR into the interactive C++ environment.

We therefore created a new, specialized library called InterOp to mitigate these problems. This library serves as a high-performance interface between C++ and Python, abstracting out details from the underlying compiler API. The InterOp library is built on top of Clang-Repl, a new capability available in LLVM, and has a reflection-focused API surface.
InterOp will follow LLVM’s release cycles, enabling its client to quickly adopt new features.

\begin{figure*}[h]
\centering
\begin{minipage}[t]{0.7\textwidth}
    \includegraphics[width=\textwidth]{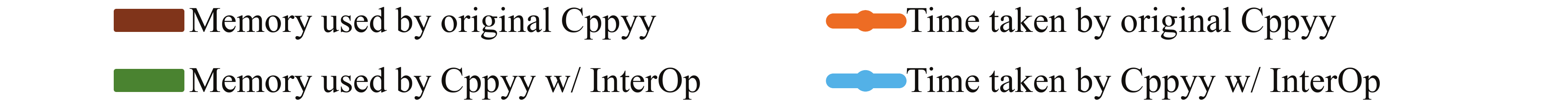}
\end{minipage}%
\quad
\begin{minipage}[t]{0.5\textwidth}
    \centering
    \includegraphics[width=\textwidth]{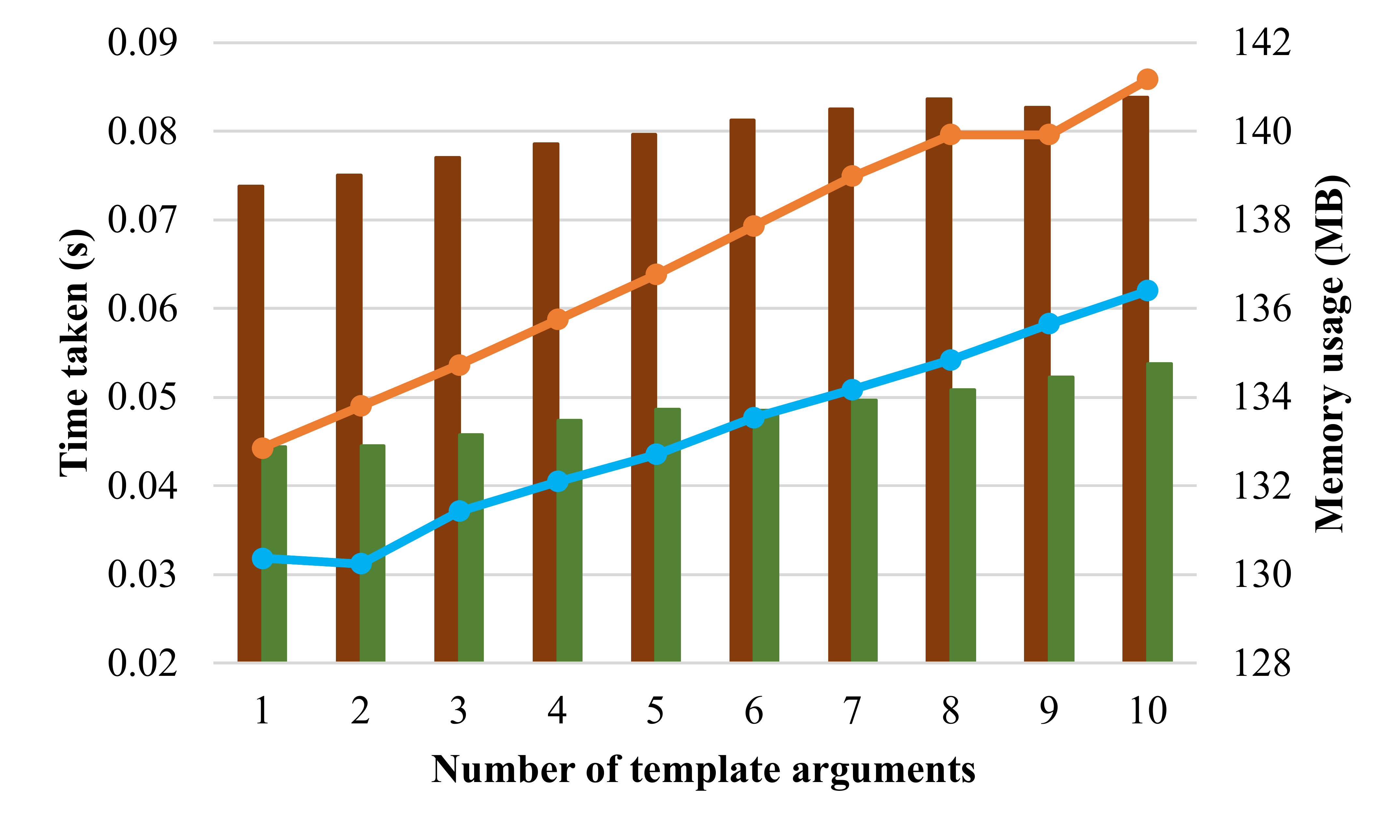}
    \label{fig:InterOpNestedTmplPlot}
\end{minipage}%
\begin{minipage}[t]{0.5\textwidth}
    \centering
    \includegraphics[width=\textwidth]{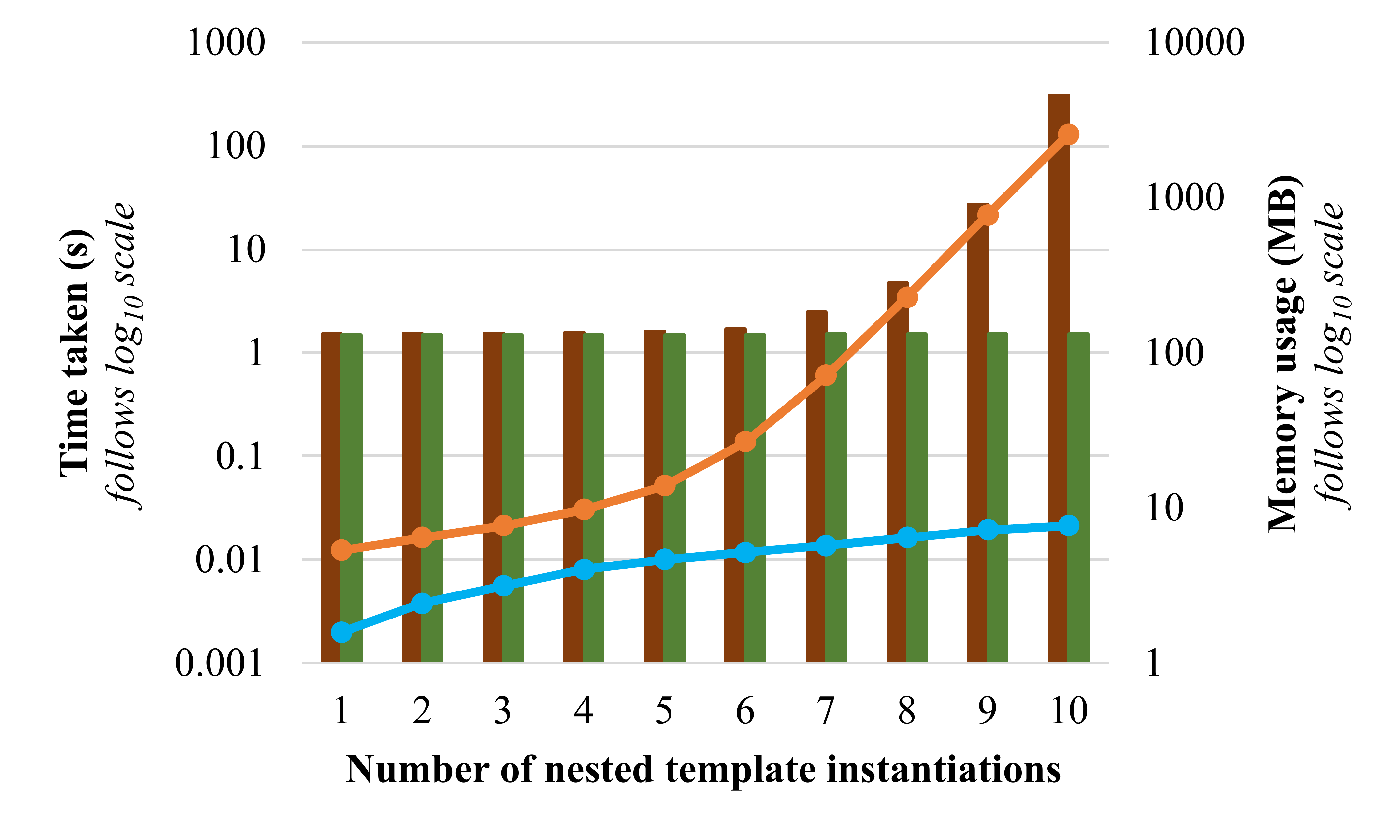}
\end{minipage}%
\caption{\textbf{Time taken and memory used during class template instantiation.} On the left, we compare template instantiations with \mintinline{cpp}{std::tuple<double, double, ...>} where the number of template instantiations done by the C++ interpreter increases with the number of template arguments. On the right, we compare instantiating nested templates, for example, \mintinline{cpp}{std::vector<...<std::vector<double> >}, where cppyy has to instantiate each nesting individually from the innermost to the outermost class template. These are common features of high-performance, templated numerics libraries that utilize template expressions.}
\label{fig:InterOpBenchmarks}
\end{figure*}

The design of InterOp brings significant improvements to cppyy in the time taken and memory used for template instantiations. It shows promising preliminary results thanks to the new way of modeling them. In Fig.~\ref{fig:InterOpBenchmarks} (left), we show improvements in unpacking templates where we recursively instantiate a textbook example of a multitype array with an underlying implementation based on \mintinline{cpp}{std::tuple}s. The InterOp version is about $40\%$ faster and $4.5\%$ more memory efficient. In Fig.~\ref{fig:InterOpBenchmarks} (right), we show that instantiating deeply nested templates scales better. The initial speedup of 6.2 times decreases to 3.8 times at about 4 levels of nesting but improves rapidly after that. In addition, we reduce the total memory used.

In addition to the improved performance, the engineering merits of using Clang-Repl and InterOp in cppyy are: (a) adopting new features faster, including potential CUDA support; (b) better runtime and memory performance due to using opaque data structures directly from the underlying compiler rather than their string representations; and (c) a wider range of supported stock LLVM versions.

\section{Related Work}
\label{sec:related}
Automatically improving the performance of Python code, i.e.\ without having to rewrite it, has been a research and/or engineering topic for many projects.
Chosen Methodologies include just-in-time compilation~\cite{PyPy}\cite{JAX}, transpilation followed by ahead-of-time compilation~\cite{ShedSkin}, or by using Python as a domain-specific language~\cite{SEJITS}\cite{Arkouda}.
Just-in-time compilation provides the most pythonic experience, however, since it can be used interactively, and is very relevant today as it is the most suitable in heterogeneous computing environments.

The PyPy project implements an alternative Python interpreter that uses a tracing JIT to improve performance.
PyPy is fully compatible with the standard CPython implementation and is able to provide significant speedups, especially for numeric code.
Cppyy is integrated at the interpreter level into PyPy, which allows C++ code to be called directly from the JITed traces without the need for function wrappers, and which enables the inlining of access to C++ data.
PyPy's tracing JIT is, contrary to all other options, not based on LLVM.
This reduces dependencies and overheads, but it does mean that the JIT can target fewer platforms and offers fewer optimizations.
A weak point of PyPy is support for existing CPython extension modules, access to which, because of different memory models, can be rather slow.
Furthermore, because the memory models need to be mapped, it is not as lenient as CPython when it comes to reference counting bugs, leading to incompatibilities.

JAX is an alternative implementation of Numpy on top of XLA, a domain-specific compiler to optimize linear algebra for machine learning.
It can JIT compile any Numpy-based code, targeting CPUs, GPUs, and TPUs.
It takes advantage of the runtime knowledge of actual data types and sizes used, to specialize the optimization.
JAX can be extended, through the underlying XLA, with user-defined functions and types.
However, there is, at the time of writing, no officially supported extension interface and suggested implementations use undocumented APIs that are unsupported and subject to change.

We find that both PyPy and JAX are good alternative candidates for our stated goals, but that Numba provides a better platform: where JAX supports GPUs well and where PyPy supports extension with user-defined functions and types, Numba does both.
We will continue to track the developments of these alternative technologies, however, and may revisit our choice in the future if warranted.

\section {Conclusion}
\label{sec:conclusion}
Seamless integration between programming languages is crucial in many scientific domains, including high-energy physics. Python and C++ are widely used in this field, and integrating the two languages is important for efficient and accessible workflows. The just-in-time compiler Numba is capable of compiling Python code to native but does not support C++. To address this issue, we introduce C++ into Numba through a new reflection interface on top of cppyy, which is an automatic, runtime bindings generator based on Cling, the C++ interpreter. This approach provides a fully automatic and transparent process, including integration with other Numba extensions, with performance on par with traditional methods. This technology allows HEP programmers to develop and debug their codes in Python, utilizing C++ libraries, while switching on the Numba JIT for selected performance-critical closures.

We demonstrate 2-20 times speedup when using Numba to accelerate cppyy through our extension. We outlined further gains using the Clang-Repl component of LLVM and the newly developed library InterOp. The preliminary results from using new infrastructure showed 1.4 to 144 times faster handling of templated code in two different scenarios in cppyy, which will indirectly improve the Numba-accelerated Python. 

\section{Acknowledgement}
This work is supported by the National Science Foundation under Grant OAC-1931408 and under Cooperative Agreement OAC-1836650.
This material is based upon work supported by the U.S. Department of Energy, Office of Science, Advanced Scientific Computing Research (ASCR), SC Program 31.

\section*{References}
\bibliography{references}

\providecommand{\newblock}{}
\begin{thebibliography}{10}
\expandafter\ifx\csname url\endcsname\relax
  \def\url#1{{\tt #1}}\fi
\expandafter\ifx\csname urlprefix\endcsname\relax\def\urlprefix{URL }\fi
\providecommand{\eprint}[2][]{\url{#2}}

\bibitem{AwkwardArray}
Pivarski J, Escott C, Smith N, Hedges M, Proffitt M, Escott C, Nandi J, Rembser
  J, bfis, benkrikler, Gray L, Davis D, Rodrigues E, Schreiner H, Gizdov K,
  Nollde and Fackeldey P 2021 scikit-hep/awkward-0.x: 0.15.5
  \urlprefix\url{https://doi.org/10.5281/zenodo.4520562}

\bibitem{Uproot}
Pivarski J, Das P, Burr C, Smirnov D, Feickert M, Gal T, Kreczko L, Smith N,
  Biederbeck N, Shadura O, Proffitt M, benkrikler, Dembinski H, Schreiner H,
  Rembser J, R M, Gu C, Edoardo, Rodrigues E, JMSchoeffmann, Rübenach J, Koch
  L, Peresano M, Eich N, Turra R and bfis 2021 scikit-hep/uproot3: 3.14.4
  \urlprefix\url{https://doi.org/10.5281/zenodo.4537826}

\bibitem{Coffea}
Gray L, Smith N, Tovar B, Novak A, Chakraborty J, Fackeldey P, Hartmann N,
  Watts G, Thain D, Stark G, BenGalewsky, Rübenach J, Fischer B, Taylor D,
  MoAly98, Kondratyev D, Gessinger P, Chen Y~M~E, Pata J, Woodard A, Potrebko
  A, Albert A, R M, slehti, Surma Z, Pedro K, dnoonan08 and Hennessee A 2023
  Coffeateam/coffea: Release v0.7.21
  \urlprefix\url{https://doi.org/10.5281/zenodo.7733568}

\bibitem{ROOT}
Brun R and Rademakers F 1997 {\em Nucl. Instrum. Meth. A\/} {\bf 389} 81--86

\bibitem{GEANT4}
Agostinelli S {\em et~al.\/} (GEANT4) 2003 {\em Nucl. Instrum. Meth. A\/} {\bf
  506} 250--303

\bibitem{Numba}
Lam S~K, stuartarchibald, Pitrou A, Florisson M, Seibert S, Markall G, esc,
  Anderson T~A, rjenc29, Leobas G, luk-f a, Collison M, Bourque J, Meurer A,
  Oliphant T~E, Riasanovsky N, Kaustubh, Wang M, densmirn, njwhite, Pronovost
  E, Totoni E, Wieser E, Seefeld S, Grecco H, Peterson P, Virshup I, G M,
  Turner-Trauring I and Bourbeau J 2022 numba/numba: Version 0.56.4
  \urlprefix\url{https://doi.org/10.5281/zenodo.7289231}

\bibitem{cppyy}
Lavrijsen W~T~L~P and Dutta A 2016 {\em Proceedings of the 6th Workshop on
  Python for High-Performance and Scientific Computing\/} PyHPC '16 (IEEE
  Press) p 27–35 ISBN 9781509052202

\bibitem{Cling}
Vassilev V, Canal P, Naumann A, Moneta L and Russo P 2012 vol 396 ({IOP}
  Publishing) p 052071
  \urlprefix\url{https://iopscience.iop.org/article/10.1088/1742-6596/396/5/052071/pdf}

\bibitem{llvmlite}
{llvmlite - A lightweight LLVM python binding for writing JIT compilers}
  \urlprefix\url{https://llvmlite.readthedocs.io/en/latest/}

\bibitem{PyPy}
{PyPy - A fast, compliant alternative implementation of Python}
  \urlprefix\url{https://www.pypy.org/}

\bibitem{JAX}
Bradbury J, Frostig R, Hawkins P, Johnson M~J, Leary C, Maclaurin D, Necula G,
  Paszke A, Vander{P}las J, Wanderman-{M}ilne S and Zhang Q 2018 {JAX}:
  composable transformations of {P}ython+{N}um{P}y programs
  \urlprefix\url{http://github.com/google/jax}

\bibitem{ShedSkin}
{Shed Skin - a restricted-Python-to-C++ compiler}
  \urlprefix\url{https://github.com/shedskin/shedskin}

\bibitem{SEJITS}
{Selective Embedded Just-In-Time Specialization}
  \urlprefix\url{https://sejits.eecs.berkeley.edu/}

\bibitem{Arkouda}
{Arkouda - Interactive Data Analytics at Supercomputing Scale a Python API
  powered by Chapel} \urlprefix\url{https://bears-r-us.github.io/arkouda/}

\end{thebibliography}

\end{document}